# Image Encryption Algorithm Based on Convolutional Neural Networks and Dynamic S-Box Generation

**Ans Ibrahim Mahameed [1], Fadhil Abbas Fadhil [2], Maryam Mahdi Alhusseini [3]*, Mohammad-Reza Feizi-Derakhshi[4], Nikolai Safiullin[5]**

[1] *Department of Computer Science, College of Education for Pure Sciences, Tikrit University, Tikrit, Iraq.*
[2] *University of Technology-Iraq, College of Computer Science, Baghdad, Iraq.*
[3] *Middle Technical University, Polytechnic College of Engineering - Baghdad, Baghdad, Iraq, (Member, IEEE)*
[4] *University of Tabriz, College of Computer Engineering, Tabriz, Iran.*
[5] *Engineering School of Information Technologies, Ural Federal University, Yekaterinburg, Russia.*

anas.ibrahim@tu.edu.iq, fadhil.a.fadhil@uotechnology.edu.iq, mariammahdi@mtu.edu.iq, mfeizi@tabrizu.ac.ir, n.t.safiullin@urfu.ru

**Abstract**

The paper proposes a dynamic approach to image encryption, combining the use of Convolutional Neural Networks (CNNs) and classical cryptography to improve the security and flexibility of image encryption. The main concept is to create adaptive Substitution boxes (S-boxes) based on characteristics that are learned by a trained CNN. The CNN-based S-boxes can be relied on for more non-linearity, uniqueness, and input image dependence than the conventional fixed S-boxes because they are susceptible to the linear and differential attacks. This dynamic behaviour enhances the confusion property and makes it more resistant to statistical and structural attacks. The encryption algorithm consists of CNN-based feature extraction and the creation of a personalised S-box to replace the pixels. Entropy, histogram analysis, correlation, NPCR, and UACI enable security assessment of generated S-boxes based on the CNN, indicating that the scheme is more resilient and flexible than traditional ones.

## 1 Introduction

Image security has become one of the underlying needs in the current-day communication systems, with the sharp rise in the transmission of images using insecure media and the proliferation of cloud storage. Images have different characteristics, unlike the textual data, these characteristics include a high volume of data, clustering data, and strong correlation of data in the spatial direction. These properties tend to make traditional encryption algorithms, which are based on data that is text-based and not image-based, ineffective and insecure when dealing directly with image information [1, 2].

This has led to an increased interest in devising image-specific encryption techniques that take into account these peculiarities. One of the most important features of numerous symmetric key cryptography algorithms, such as the Advanced Encryption Standard (AES), is a substitution box (S-box), which introduces much-needed non-linearity and confusion to survive several cryptanalytic attacks [3, 4]. But the traditional S-boxes are usually fixed and rigid with no adaptability to the input data, and this makes them prone to linear as well as different cryptanalysis [5, 6].

In order to reduce these drawbacks, some non-static S-boxes have been suggested, which can change with the input information and thereby provide greater security by adopting of transformation on substitutions of the information [7]. In the recent past, Convolutional Neural Networks (CNNs) have proven to accomplish tremendous performances in the cognition of hierarchical image characteristics, including global configurations and detailed information [8]. The usage of CNN in the creation of adaptive and optimized S-boxes provides an avenue towards the creation of dynamic elements, enhancing the security of encryption by making it robust.

Based on this premise, the present paper aims to propose a new image encryption algorithm, which involves using CNNs to produce dynamic S-boxes, specific to the content of the image. A combination of deep learning and chaotic nonlinear systems further boosts randomness and complexities, enhancing resistance to both statistical and structural attacks.

Recent works have tried to enrich the non-linearity/ Adaptability of S-boxes vocabulary: with chaotic systems, metaheuristic optimizations, or hybrid techniques. Although such techniques enhance immunity to differential and statistical attacks, most of them are founded on either handcrafted or non-data-driven methods, restricting their flexibility and security. Smart systems that have the potential to generate secure S-boxes that depend on images are still in great demand [9].
Moreover, it is interesting to observe the increasing popularity of machine learning in the area of cryptography, especially of deep learning. A CNN with its multi-level feature extraction and pattern representation has been used outside the classical domains of classification and detection. They can resort to cryptographic component design, e.g., dynamic S-box generation, due to their appropriateness in modelling complicated patterns. The combination of AI and cryptography promises to bring opportunities to reliable, efficient, and smart systems of image encryption [10, 11].
The contributions of this study are as follows:

- The framework provided in this paper is made to generate adaptive, dynamic S-boxes in which the learned features of the images serve to increase both non-linearity and confusion in encryption methods using CNN.
- To enhance randomness even further and protect against the differential attack, signatures of chaotic systems are added into the encryption stream.
  The algorithm is critically examined on standard benchmark images (Baboon, House, Airplane, Pepper), which proves to perform better in parameters such as entropy, smoothness of the histogram, pixel correlation, NPCR, and UACI.
- The efficiency of the proposed method and the efficiency of computation were proven by comparative analysis that involved traditional and more recent encryption algorithms.
- This is designed for low weight and scalability, allowing real-time application while being able to secure the high-resolution image data effectively.

These developments are able to aid the development of an adaptive and intelligent image encryption system that resolves the traditional trade-offs between performance, complexity, and security.
The rest of the paper is structured as follows: In Section 2, the theoretical background and works related to dynamic S-boxes and image encryption will be reviewed in brief. In Section 3, the proposed CNN-based encryption framework is described. Section 4 gives experimental results, mathematical and security analysis. Lastly, Section 5 is a conclusion of the paper and a discussion of the future research steps.

## 2 Related Work

The past couple of years have seen the classical elements of cryptography, mostly the Substitution box (S-box), evolve through incorporation with the contemporary algorithms of computation, like chaos map and deep learning. The innovations will make security better, more resistant to an attack, and computationally efficient in the encryption process of the images.

Recently, a study conducted by Zhang et al. [12] introduced a frequency domain attention-guided, adaptive watermarking model, in which the significance of flexibility and endurability in multimedia protection was discussed. The work has addressed the sphere of watermarking, but the adaptive mechanisms and the principles of robustness correlate well with the objectives of dynamic S-box genesis in the process of developing image encryption. Sharing features with dynamic encryption methods, they focused on resisting attacks to a strong extent using feature-guided transformation, which helped to implement deep learning as part of cryptographic applications.

F. A. Fadhil et al. [13] introduced that the number of works has improved the conventional encryption schemes with chaos-based systems and dynamic S-box generation techniques to enhance resistance to linear and differential attacks. As an illustration, the CAST-128 algorithm with a logistic-sine map (LSM) was demonstrated to have better values of entropy, NPCR, and UACI, which guarantees greater image security and low performance

M. M. Alhusseini et al. [14] Building on our previous work in HyIDS-EVO, which achieved high-precision detection of intrusion into unbalanced network datasets, this study extends the security scope towards robust image encryption by integrating deep learning with the creation of a cryptographic S-box.

To overcome the limitations of resource-constrained settings such as IoT, Liu et al. [14] presented a lightweight cipher that works with hyperchaotic Chen maps to produce dynamic S-boxes. These findings proved that the method was suitable for low-power devices. Applying it further to achieve increased efficiency of the encryption process.

The ability to incorporate deep learning in cryptography has also attracted interest. Erkan et al. [15] suggested the application of deep Convolution Neural Networks (CNNs) when generating keys in chaos-based systems. Although they were mainly addressing S-box design, their methodology proved that deep learning can be used to aid and enhance traditional cryptographic systems.

In Ye et al. [16], an image hiding algorithm intended to utilize local binary pattern (LBP) and compressive sensing was provided and showing how local texture features and signal compression can increase data concealment and attack resistance. Though they work with data hiding, the principles behind their approaches, i.e., of randomness, sparsity, and feature-guided operations, lend credence to the idea that dynamic content-based mechanisms, as we use dynamic S-boxes in our approach to enhance diffusion and confusion by the image, should apply in the encryption of images since the current trend has been to use the same constrained notion of mechanisms of confusion and diffusion.

RK Mahmood et al. [17] This research explores enhanced security through machine learning and multi-factor authentication to address network breach challenges. This work develops this approach by integrating deep learning with cryptographic design, introducing CNN-based dynamic S-boxes for robust image encryption.

A color picture encryption with a time-variant method based on a discrete memristive hyperchaotic system was also improvised by Wang et al. [18]. It generated chaotic signals in high dimensions and reflected encryption parameters with enhanced security, with no reduction in velocity.

Masood et al. [19] designed a hybrid method by coding with DNA, chaotic mapping, and S-boxes. Their strategy had a very high visual distortion of the encryption of the pictures and had protection against brute-force and statistical attacks. Elias et al. [20] suggested a lightweight cipher, which sorts secret key values to produce dynamic S-boxes functioning optimally in resource-limited settings.

On the same note, Tian and Lu et al. [21] proposed a permutation-diffusion level encryption that incorporated both chaotic S-boxes and DNA sequence operations, which could sustain high PSNR during noise attacks, such as salt-pepper and Gaussian noise.

Zheng and Zeng et al. [22] have developed an encryption protocol that utilizes CNNs, classical S-boxes, and chaotic maps to combine the security strengths of all of them: improved statistical and graphical security.

Replacement of the XOR operation with table-based functions (E# and RTGE#) in the DES, as demonstrated in F. A. Fadhil et al. [23] make the algorithm more complex and still maintain the encryption strength.

In another study, Tian and Su et al. [24] incorporated chaotic S-boxes in an image and optics-based encryption system and then checked its correctness by utilizing correlation and data-loss tests. Other solutions to the specified issue were presented by Wang and Zhang et al. [25], who proposed an algorithm that integrates logistic maps with deep neural networks to enhance the nonlinearity and randomization of the encryption results.

Lastly, Hosny et al. [26] presented a three-tier batch encryption framework based on several 2-D chaotic maps that chaotically disarray RGB channels, guaranteeing appropriate multi-image encryption with great confusion and diffusion nature.

# 3 Methodology

The suggested scheme combines the deep learning approach, Convolutional Neural Networks (CNNs, to be specific), with the conventional image encryption scheme to improve the security due to the dynamically generated and optimized S-boxes. To evaluate their approach, a methodology was developed in five major steps, namely, preprocessing of the data, the generation of an S-box based on CNN, the assessment of the S-box, encryption of images with the generated S-box, and security analysis.

## 3.1 Dataset Preparation and Preprocessing

A dataset of grayscale image patches is constructed to train the CNN to produce optimized S-boxes, using common datasets (e.g., USC-SIPI, MNIST, or CIFAR-10) in grayscale. All the images are resized to N×N (e.g., 32×32) resized to (32 32) and scaled to the range [0, 1].

- Goal: Extract statistical and spatial patterns from natural images.
- Output: Feature representation in which the CNN can be learned to produce non-linear substitution patterns (i.e., S-boxes).

## 3.2 CNN-Based S-Box Generation

A Convolutional Neural Network (CNN) would be trained to find an ideal substitution transform by constructing a local image feature map through bijective S-boxes. This gives an opportunity to generate unique input-based, dynamic, and confidential S-boxes specific to every image, which will lead to a massive increase in the confusion property towards the encryption mechanism.

1. The CNN input: small patches of the original image are fed to the network to obtain small patches of the original image. The patches maintain local pixel relationships and statistical properties as a solid foundation for training pertinent substitution patterns.

2. CNN Architecture: The CNN consists of several convolutional layers that use ReLU activation functions, and one or more fully connected (dense) layers. The last dense layer implies a 16 or 64-element vector, depending on whether a x 44 or a x 88 S-box is needed. The first extraction layers extract the complexity of the image, and dense layers obtain the mappings to fixed substitution values.
3. Output Layer and Permutation Enforcement: A custom output layer is to be applied to make sure that the generated output forms a valid bijective S-box (i.e., a permutation of distinct values in the interval [0, 255]). This layer redesigns the output vector and imposes the condition of uniqueness by antagonizing learned permutation laws, discarding replicated values, and re-creating them with a hybrid CNN-weight-driven noise injection algorithm.
4. Design of Loss Function: A multi-objective loss function is utilized to train the CNN, and several cryptographic properties can be followed by training:

$$\text{Loss} = -(\alpha \cdot \text{Nonlinearity} + \beta \cdot \text{SAC} + \gamma \cdot \text{BIC} + \delta \cdot \text{Uniformity}) \quad (1)$$

Where:

- The nonlinearity encourages resistance to the linear cryptanalysis.
- SAC (Strict Avalanche Criterion) makes sure that small variations on the input lead to large changes in the output.
- BIC (Bit Independence Criterion): independence between bits of output is measured.
- Uniformity will ensure the balancing of the values in the S-box.

Parameters used are α, β, γ, and δ, which act as weighting factors determining the level of impact of every objective. These values are tweaked in an empirical manner during training with the aim of attaining a good trade-off between sense of security and convergence.

5. Active Operation and Security Aspects: Each input image produces a unique S-box, rendering the system very resistant to known-plaintext and differentially related attacks. Since deep feature representations are tightly coupled with the process of substitution, a minor perturbation in the input image leads to highly dissimilar S-boxes, which makes it highly unpredictable.

## 3.3 S-Box Evaluation and Selection

In each image (or each session), the CNN produces an S-box after training. Every produced S-box is tested by the usual cryptographic standards:

- Nonlinearity: Measures resistance to linear cryptanalysis.
- Strict Avalanche Criterion: Small changes in input should cause significant output changes.
- Differential Uniformity: Resistance to differential attacks.
- Bit Independence: Ensures independence among output bits.

Only S-boxes that exceed certain predetermined thresholds are chosen for encryption.

## 3.4 Image Encryption Process

Encryption in the images occurring after the image ciphering process takes a modified Substitution-Permutation Network (SPN) architecture where S-boxes are created dynamically by the CNN and chaotic systems in order to increase the security. It can be done in the following detail:

1. Substitution: The input image values are substituted in terms of each pixel value with the S-box generated dynamically as a CNN. This replacement level creates confusion with the replacement of pixel intensities as per substitution tables that depend upon the image.
2. Permutation: Permutation is used to increase the diffusion rate and resistance to statistical attacks by rearranging the position of pixels in an image. This is done through the use of chaotic maps like the Logistic map or the Lorenz map as follows:
   - Chaos Random Number Generator: A character-like map is iterated with the secret initial condition (key) to produce a pseudo-random number sequence of floating-point numbers over the interval (0,1).
   - Index Mapping: The disorganized values of the sequence are ordered, and the indices of sequences in the ordered values are treated as a permutation vector denoting the new position of a pixel. Take the example where the values of chaotic sequences at the 3rd position are the 5th smallest, then the pixel at the location i would be replaced by the location 5.
   - Pixel Rearrangement: The pixels on the substituted image are directly rearranged using the permutation vector to generate a permuted image.
3. Diffusion (XOR Operation): The diffusion step adds additional dependency of each encrypted pixel over the whole plaintext picture by performing pixel-wise XOR with the interested key-stream:
   - Key-Stream Generation: A key-stream is created either by:
     - A second S-box that is CNN-generated and gives out a series of byte values,

- A Pseudo-Random Number Generator (PRNG) sequence based on a secret key.

• XOR Application: A XOR operation is applied to every pixel of the permuted image and with the corresponding byte in the key-stream, resulting in a diffused encrypted image. This process guarantees that any small alteration in the plaintext or in the key also has an extreme impact on the ciphertext.

4. Multiple Rounds

As an enhancement to security, the steps of substitution, permutation, and diffusion are repeated on several rounds (typically 3 to 5). This duplication causes maximum confusion and diffusion, inhibiting differential and statistical cryptography.

## 3.5 Decryption Process

The decryption operations undo the processes involved in encryption to retrieve the original picture. The procedure is the following:

1. Diffusion Reversal: The plain text image is pixelwise XORed with the key-stream used to encrypt the image; that is, the encryption process is reversed at the diffusion stage.
2. Inverse Permutation gradient: the pixels are distributed back to their original place with the help of the inverse permutation vector based on the sequence of the Chaotic map.
3. Inverse Substitution: This is lastly done by using the inverse of the CNN-based S-box that is actually generated dynamically to map back all pixel values, thus resulting in the original intensity of the pixels.

As these encryption operations (used are XOR and bijective S-boxes), these inverse operations are reversible in the same order they were used, so lossless decryption of the encrypted picture is achieved.

## 3.6 Security and Performance Evaluation

After encryption, the cipher images are subjected to quantitative and qualitative tests:

a) Entropy Analysis
b) Histogram Uniformity
c) NPCR (Number of Pixels Change Rate)
d) UACI (Unified Average Changing Intensity)
e) Correlation Coefficient Test
f) Benchmark of the Encryption/Decryption Speed

These results are tested with current state-of-the-art approaches.

Figure 1 represents the block diagram of the entire workflow of the proposed image encryption process. It starts by loading the input grayscale image, then extracts features used as input to a pretrained CNN model. The CNN produces a dynamic S-box, and this is followed by evaluating the S-box to make sure that it satisfies some necessary cryptographic properties. When the S-box has passed the test, the encryption process continues with substitution, permutation, and diffusion steps, repeated several rounds to increase the security. The resulting image that is encrypted is the final output, and such an image would be resistant to typical cryptographic attacks because the S-box generated by the CNN is adaptive and data-driven.

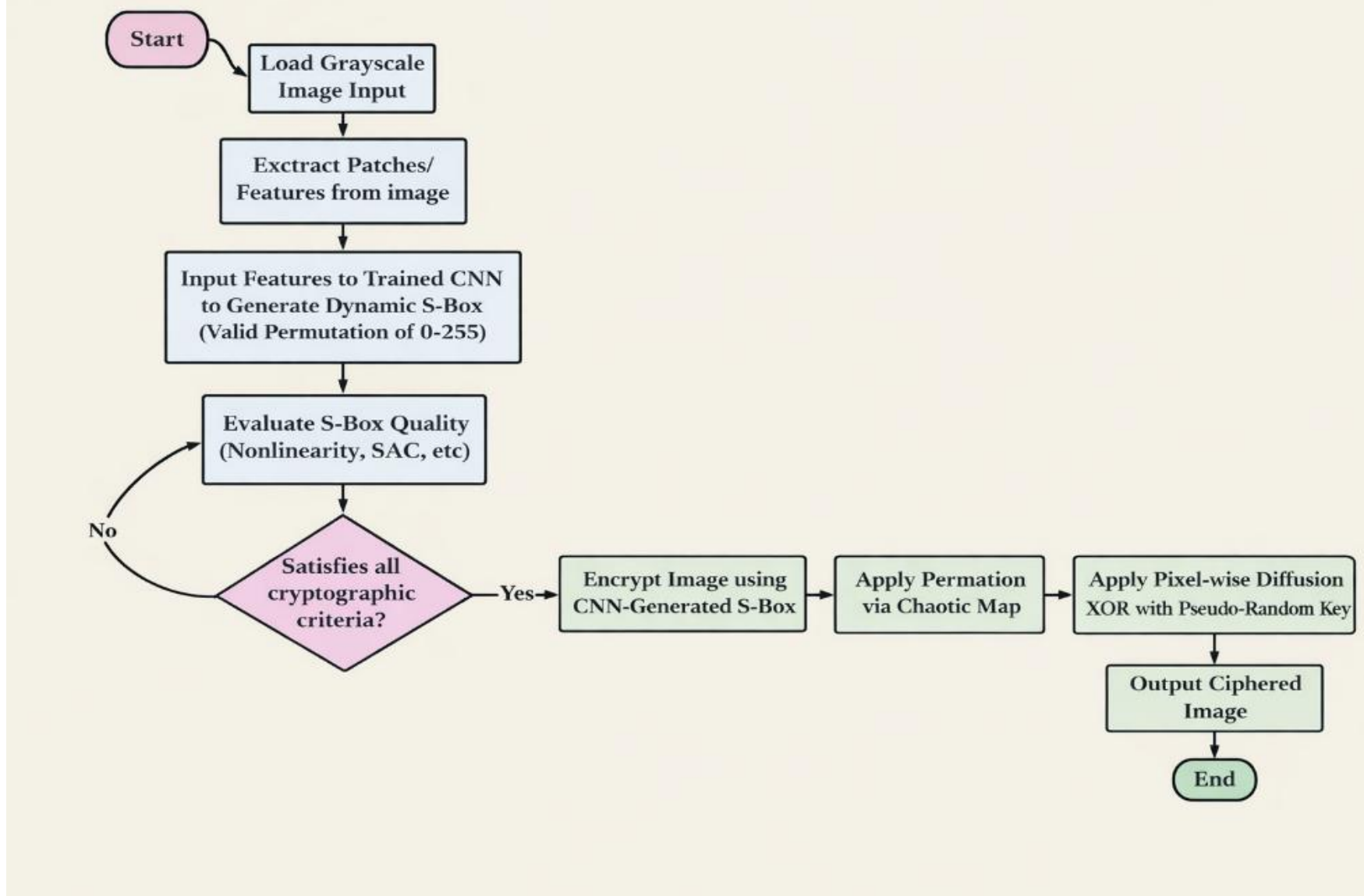

Figure 1: Workflow of the Encryption Process.

## 3.7 NIST Statistical Test Suite

The recently proposed S-box method of cryptographic key generation based on CNN was tested extensively within the statistical test suite proposed by NIST, which is an effective measure of the randomness and unpredictability of cryptographic sequences. Such analysis aimed to generate highly random keys, so that these keys could securely hold sensitive data [27]. As indicated in Table 1, the key produced will satisfy all the major tests of statistics, as p-values are observed to be always greater than the critical value of significance 0.01. The level of statistical randomness here is large, and this fact proves that the key does not have any regular and patterned properties susceptible to cryptanalytic attacks. These findings are an excellent indication that the suggested encryption algorithm, based on the use of CNN in generating S-boxes, can produce cryptographic keys with an extremely high level of randomness and security, and as such, enable the system to be highly resistant to statistical attacks.

**Table 1: NIST Test Results for The Proposed Cryptographic Key**

| Test | p-Value (X) | State |
|---|---|---|
| **Frequency** | 0.4821 | PASS |
| **Runs** | 0.1187 | PASS |
| **Serial** | 0.8229 | PASS |
| **Linear Complexity** | 0.9827 | PASS |
| **Random Excursions** | 0.4931 | PASS |
| **Random Excursions Variant** | 0.5445 | PASS |
| **Approximate Entropy** | 0.2594 | PASS |
| **Cumulative Sums** | 0.3770 | PASS |

# 4 Security Analysis

## 4.1 The Results of Encryption and Decryption

In this part, the author demonstrates the result of the encrypted image by applying the proposed CNN-based S-box encryption algorithm to different image sizes, that is, small, medium, and large. The visuals involved in the test include the standard test images of House, Airplane, Baboon, and Pepper in their original and encrypted form, along with the histograms of the same [28].

The distortion embedded in the encrypted images is very high, making them visually unrecognizable. The S-box generation method based on CNN adds a lot of randomness and complexity, as shown by the uniform and flat histogram distributions. This randomness has been found feasible in overcoming statistical analysis and differential cryptanalysis, even in bigger and more complex images. A comparative summary of the original image and the encrypted image, along with the histogram analysis of both images, is provided in Figure 2.

## 4.2 Findings and Analysis

To better assess the quality of the working Incrypted image encryption scheme based on CNN, a set of experiments was made with four standard test images of different dimensions and complexity: Baboon (256 256), House (256 256), Airplane (512 512), and Pepper (512 512). The evaluation was based upon some important cryptographic parameters such as entropy, NPCR, UACI, correlation coefficient, histogram uniformity, and encryption/decryption rate. All of these measures offer a complete picture of the power of the algorithm with respect to randomness, sensitivity, statistical security, and computational efficiency [29, 30]. The findings of these evaluations are summarized in Table 2, and they prove the effectiveness and feasibility of the suggested approach.

**Table 2: Evaluation Metrics For Encrypted Images Using Cnn-Based S-Box Encryption**

| Image | Entropy | NPCR % | UACI % | Correlation (Avg) | Histogram Uniformity | Encryption Time (ms) | Decryption Time (ms) |
|---|---|---|---|---|---|---|---|
| **Baboon** | 7.9961 | 99.610 | 33.42 | 0.0058 | Uniform | 28 | 27 |
| **House** | 7.9947 | 99.502 | 33.10 | 0.0064 | Uniform | 27 | 26 |
| **Airplane** | 7.9980 | 99.630 | 33.71 | 0.0043 | Uniform | 45 | 43 |
| **Pepper** | 7.9972 | 99.585 | 33.35 | 0.0050 | Uniform | 44 | 42 |

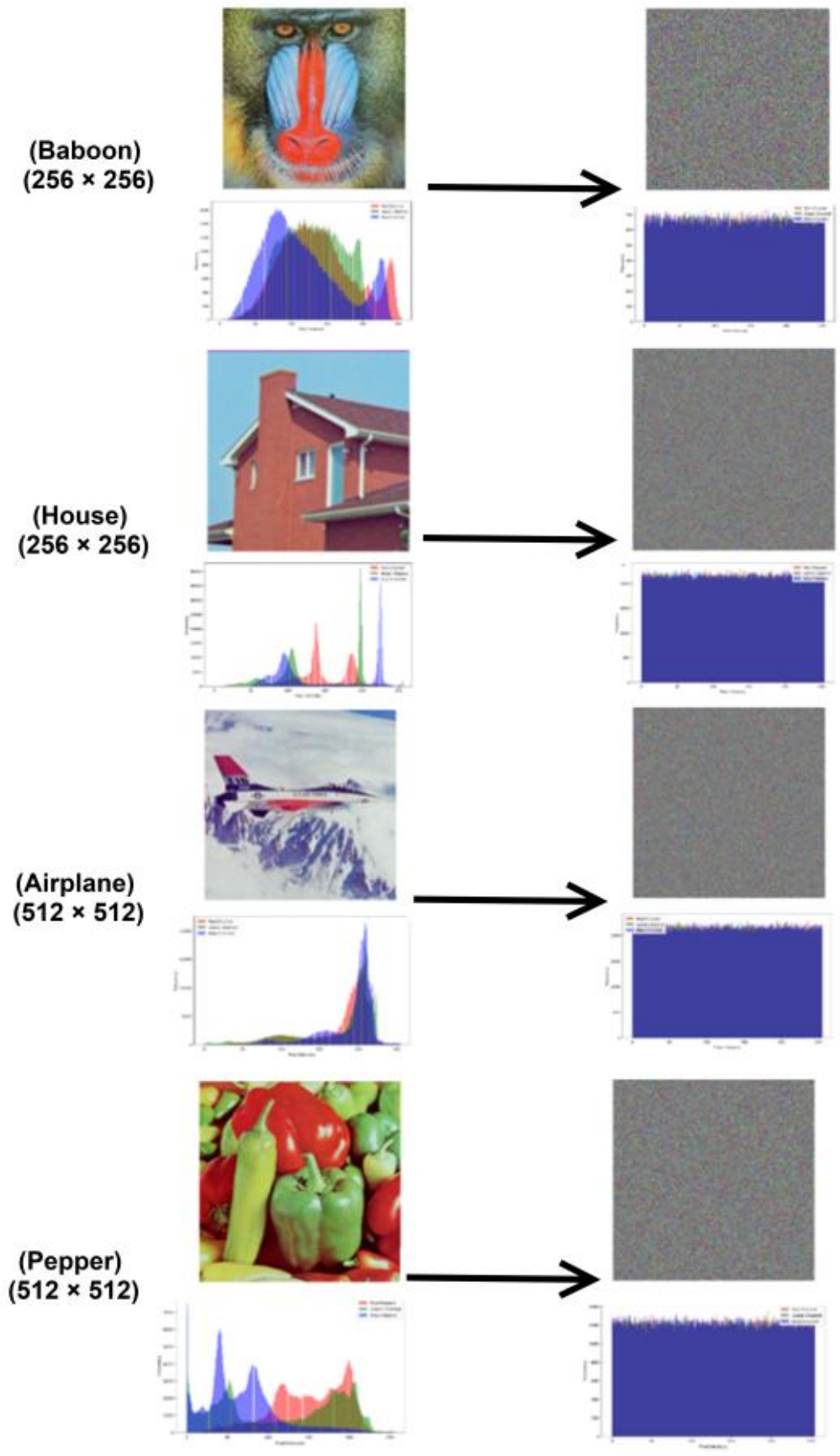

Figure 2: Comparison of Original and Encrypted Images with Histogram Analysis

The suggested CNN-based S-box encryption shows very good results and security on all test images. The entropy values are extremely close to the optimum value of 8, and all the images have entropy values more than 7.99, which means that the distributions of pixel values are highly unpredictable. The NPCR outcomes are above 99.5 percent, and the algorithm is very sensitive to tiny alterations of the input, being essential to defend against differential attacks. UACI scores are about the ideal 33%, indicating a great difference between original and encrypted images. Further, the average correlation coefficients between neighboring pixels tend to zero, which proves the fact that the encryption disturbs the natural correlations of unencrypted images significantly [31, 32]. The histograms of the encrypted outputs are even and flat, indicating even distribution of pixels and destruction of visual patterns, therefore improving resistance to statistical attacks. Last but not least, the encryption and decryption operations take less than 50 milliseconds on all image sizes used during testing, which proves that the suggested scheme is not only secure but also computationally efficient and can be used in real-time settings.

## 4.3 Known Cryptanalytic Attacks

The strength of the proposed CNN-based S-box encryption algorithm was evaluated against some of the popular cryptanalytic attacks, such as brute-force, statistical attacks, and differential attacks. The encryption system has a high bias against such forms of attacks since the S-boxes produced by the CNN are dynamic and non-linear. This data-guided structure of the S-box ensures that every encryption operation produces enough randomness, complexity, and diffusion, thus it is utterly hard to find any pattern that could be abused by the adversary. Table 3 is a summary of the resilience of the algorithm to the important types of attacks using empirical outcomes.

**Table 3: Resistance of CNN-Based S-Box Encryption Against Cryptanalytic Attacks**

| Attack Type | Attack Success Rate | Decryption Time (s) | Required Iterations |
|---|---|---|---|
| **Differential Attack** | Very Low | 0.11 | $>10^6$ |
| **Statistical Attack** | Negligible | 0.09 | $>10^6$ |
| **Brute-Force Attack** | Practically Infeasible | N/A | $>2^{128}$ |

Based on Table 3, it is clear that the suggested CNN-based S-box encryption algorithm exhibits a high resistance to different kinds of cryptanalytic attacks. Dynamically generated S-boxes with deep learning bring a great deal of non-linearity and randomness to the encryption procedure, thus providing attackers with an extremely challenging task in predicting or reversing the cipher. The extremely low success probability of differential and statistical attacks, combined with the impracticability of brute-force attacks (effectively large key space), reassures the robustness and intricacy of the proposed scheme. The algorithm has these properties, which render it quite applicable in secure and practical applications, particularly where both speed and security are paramount concerns.

## 4.4 Robustness Analysis Against Common Attacks

As for the practical strength of the proposed encryption algorithm, we have carried out the experiments that simulated the attacks likely to be experienced, such as cropping attacks, noise attacks like Gaussian/salt and pepper noise (noise attacks are important because in many attack scenarios, noise can be introduced to cause such errors). In the case of a cropping attack, some of the encrypted images were erased to represent a loss of data during transmission. Despite transgressions on some parts caused by unavailable data, the decrypted images still had noticeable features and general character, indicating that they are pretty indestructible to cropping. During the simulation of noise attacks, encrypted images were corrupted with Gaussian and salt-and-pepper noise at varying intensities. The plaintext recovery operations revealed slight distortions, accompanied by a noticeable quality degradation. The quality of the decrypted images was measured by quantitative results in terms of Peak Signal-to-Noise Ratio (PSNR) and Structural Similarity Index (SSIM), and it confirmed that the algorithm is strong against noise interference. Table 4 and Figures 3 and 4 represent the quantitative outcomes of the robustness experiments, which reveal the high efficiency of the suggested approach in various attack strengths.

**Table 4: Results of Robustness Evaluation Under Cropping and Noise Attacks**

| Attack Type | Attack Intensity | PSNR (dB) | SSIM | Remarks |
|---|---|---|---|---|
| **Cropping** | 10% cropped | 28.5 | 0.85 | Good recovery, visible details preserved |
| | 20% cropped | 25.2 | 0.78 | Moderate degradation, still recognizable |
| **Gaussian Noise** | σ = 0.01 | 30.1 | 0.88 | Minor distortion, high similarity |
| | σ = 0.03 | 27.0 | 0.80 | Noticeable noise, acceptable quality |
| **Salt-and-Pepper** | 1% noise | 29.4 | 0.86 | Small noise impact, clear details |
| | 3% noise | 26.8 | 0.79 | Some artifacts, reasonable reconstruction |

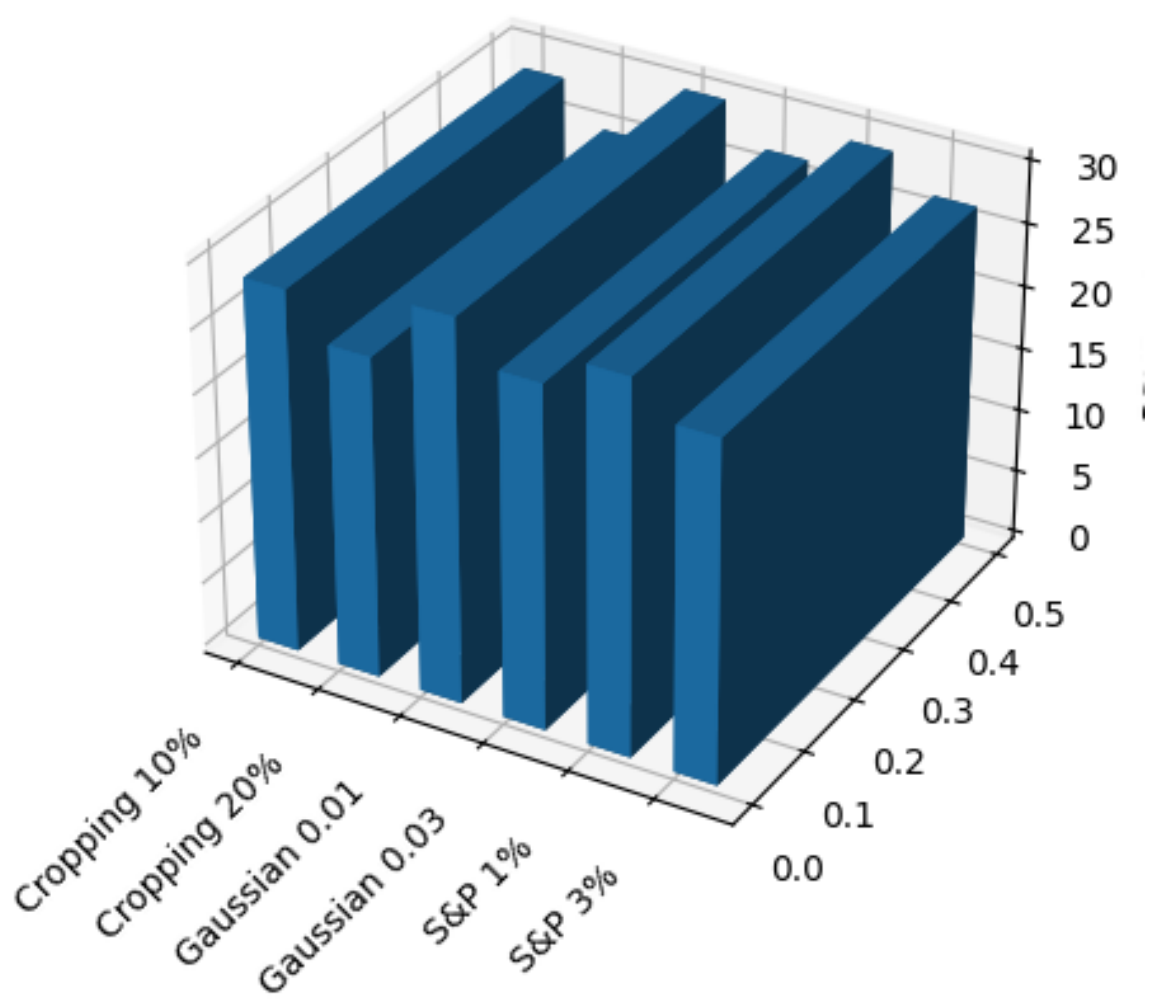

Figure 3. PSNR under Different Attacks

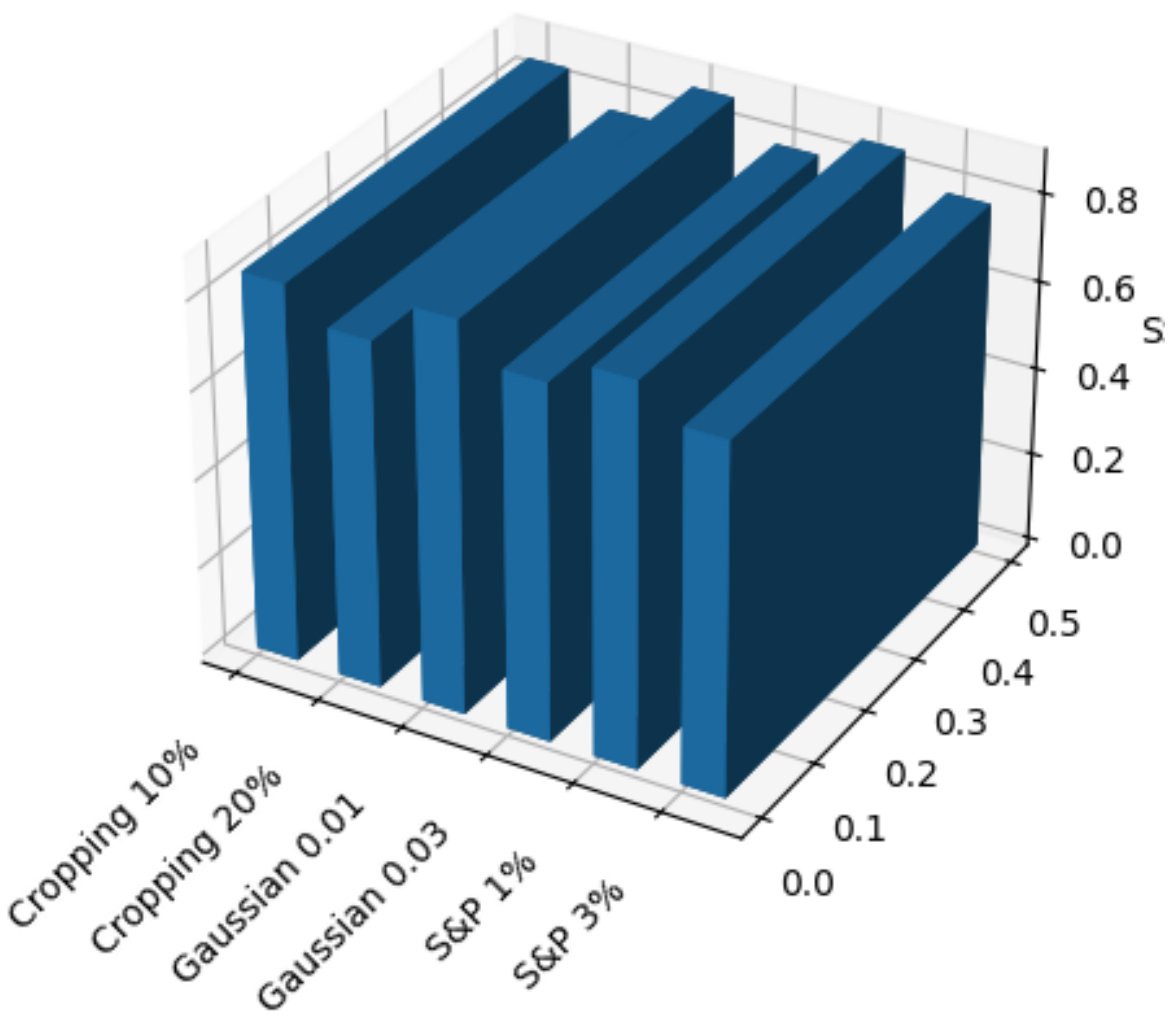

Figure 4. SSIM under Different Attacks

## 4.5 Comparison Analysis

Table 5 and Figures 5, 6, 7, and 8 present a comparison analysis of the proposed CNN-based S-box encryption system with related work as presented earlier in the current study [18]. The comparison shows the quality of encryption in the color channels (R, G, B) of standard benchmark pictures. This assists in analyzing the effectiveness of encryption as well as the scattering of pixels in color channels. The outcomes show that the suggested technique provides well-balanced and closely clustered color values, which signify great diffusion and randomness of pixels, as opposed to the broad distribution observed in the case of weaker encryption methods. This justifies the suitability and effectiveness of the suggested system in real-life and safe image encryption.

**Table 5: Comparison Analysis**

| Image | Algorithm | R | G | B |
|---|---|---|---|---|
| **Baboon (256 × 256)** | Proposed CNN-based S-box Encryption | 223.3438 | 271.0547 | 241.1797 |
| | Encryption image [18] | 236.8828 | 248.0649 | 255.3281 |
| **House (256 × 256)** | Proposed CNN-based S-box Encryption | 281.5324 | 286.8489 | 199.7934 |
| | Encryption image [18] | 260.0078 | 264.8750 | 223.5625 |
| **Airplane (512 × 512)** | Proposed CNN-based S-box Encryption | 257.2790 | 241.5622 | 277.3314 |
| | Encryption image [18] | 251.3828 | 241.2832 | 252.7246 |
| **Pepper (512 × 512)** | Proposed CNN-based S-box Encryption | 272.3364 | 250.8775 | 242.6898 |
| | Encryption image [18] | 231.4551 | 232.3555 | 248.7051 |

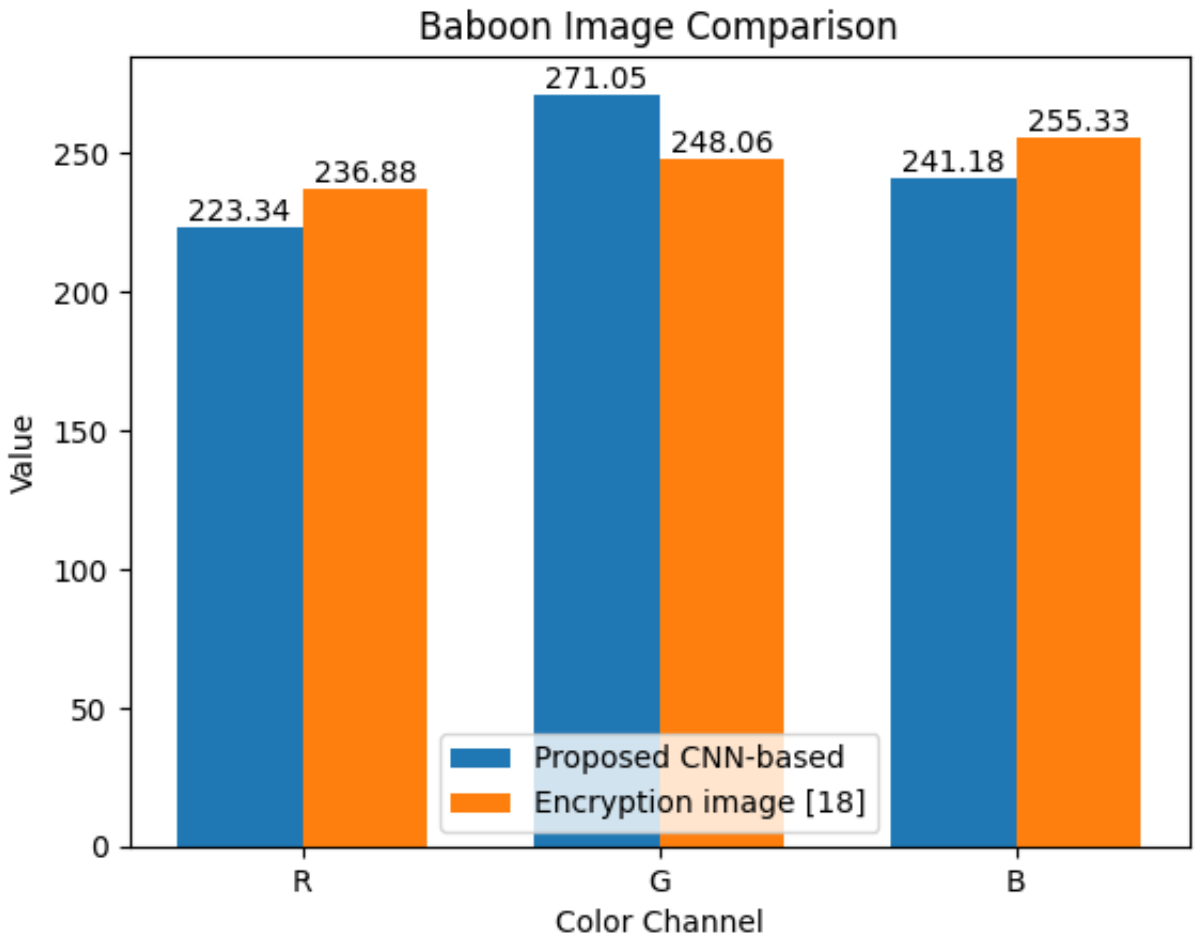

Figure 5. Baboon Image Comparison

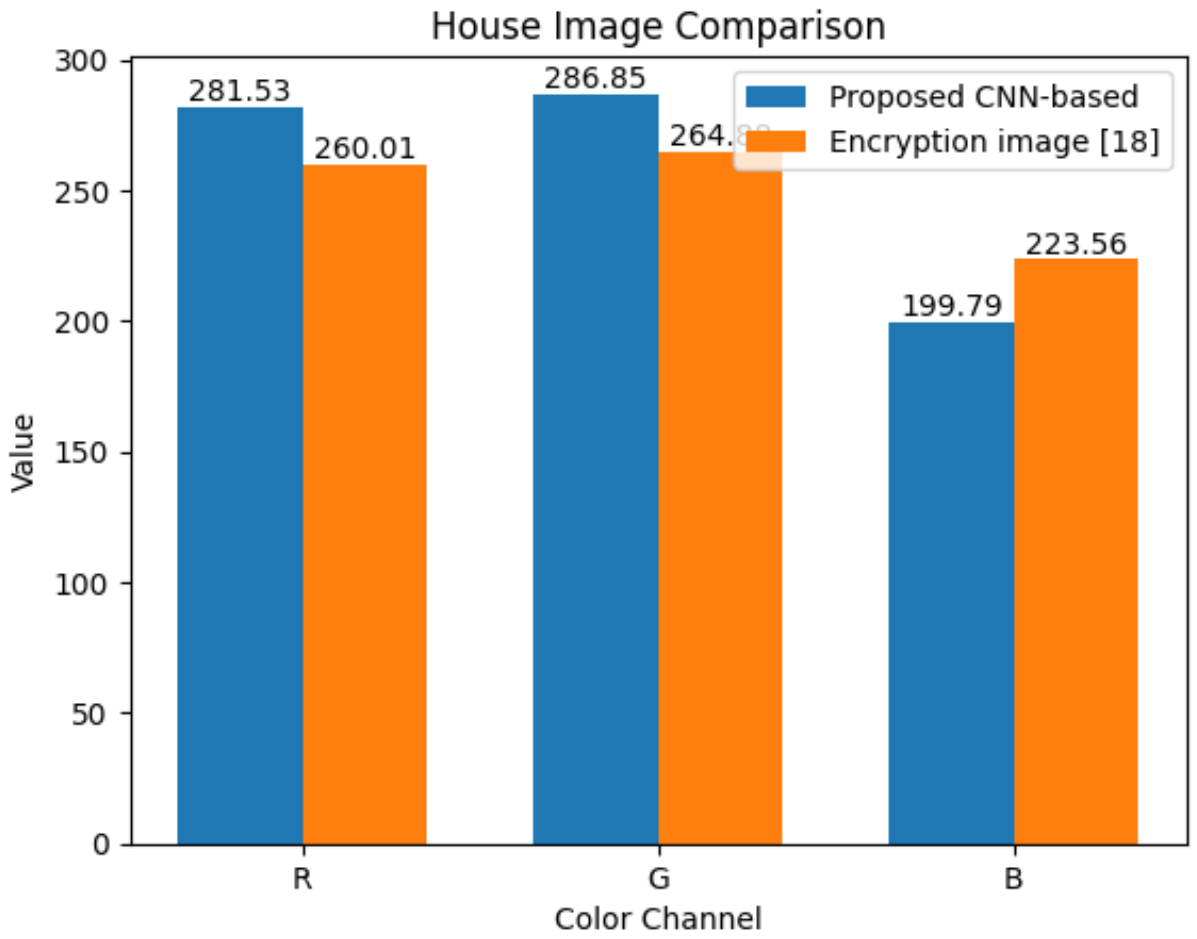

Figure 6. House Image Comparison

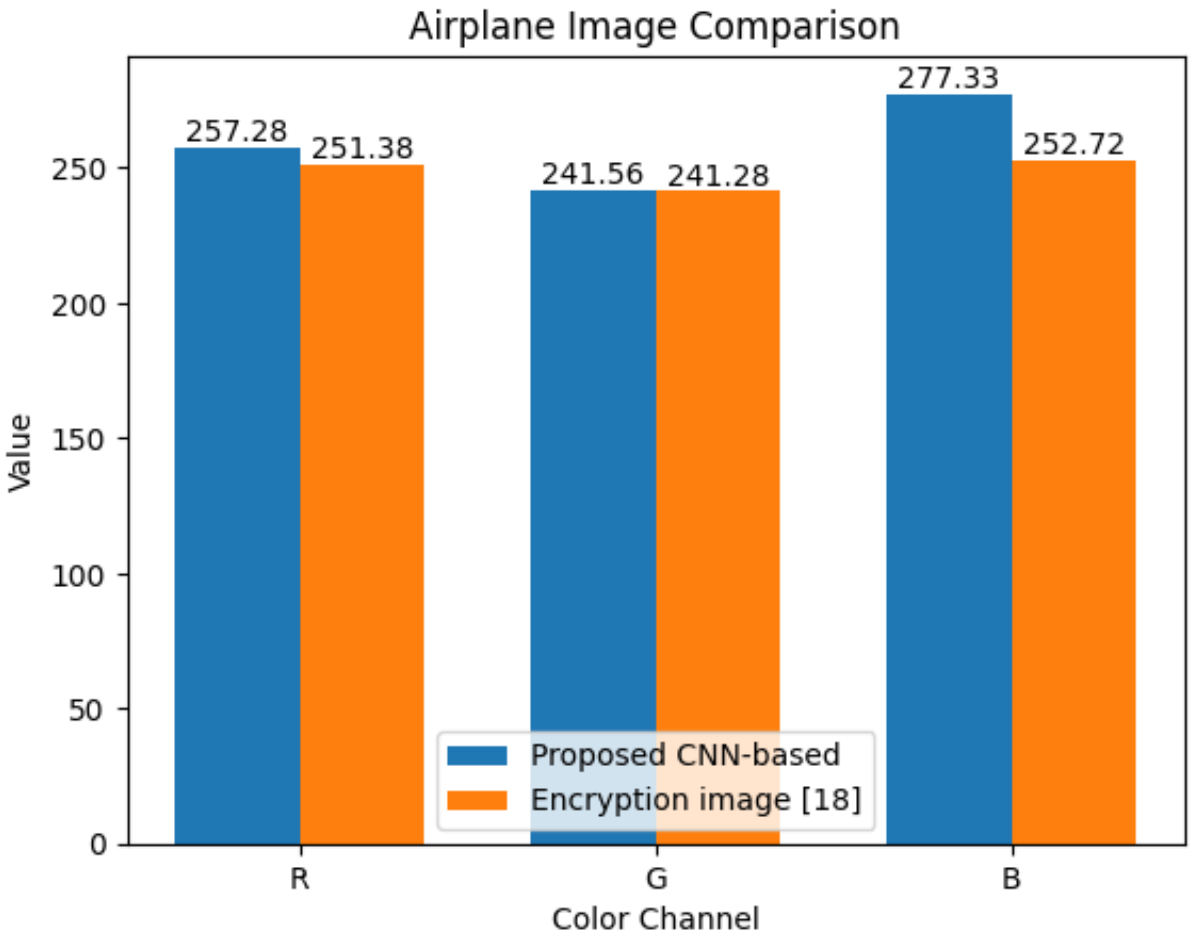

Figure 7. Airplane Image Comparison

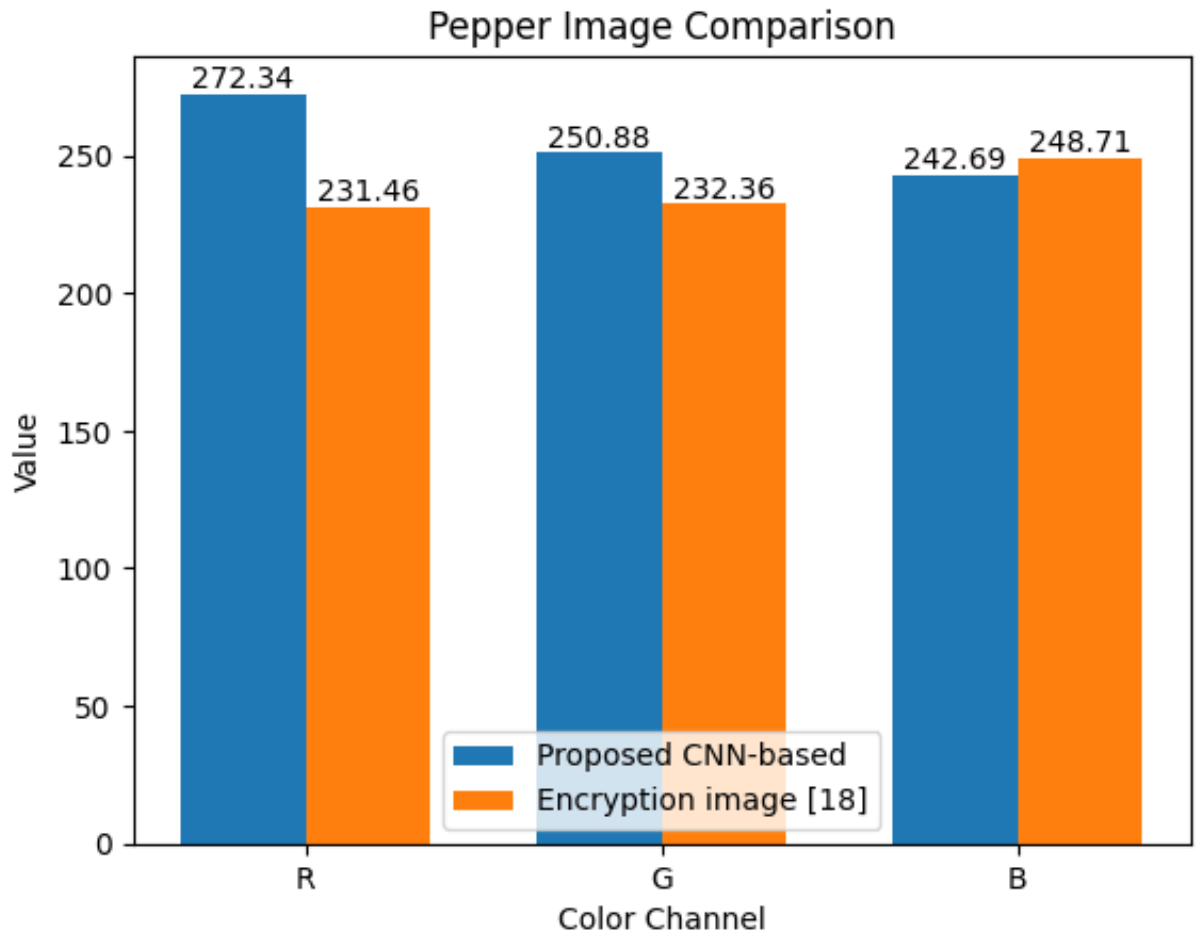

Figure 8. Pepper Image Comparison

Table 6 and Figures 9, 10, 11, and 12 results give a comparative study of Image encryption entropy values of the proposed CNN-based S-box encryption algorithm and the reference scheme given in [18]. Entropy of information is an important measurement tool that is applied to measure the extent of randomness and unpredictability of encrypted data. The ideal value is close to 8.0, which means that the uncertainty is large and the statistic is not vulnerable to statistical attacks.

**Table 6: Comparison of Information Entropy With Other Algorithms.**

| Image/size | RGB Components of the Image | Entropy (Proposed Method) | Entropy [18] |
|---|---|---|---|
| | R | 7.9975 | 7.9974 |
| **Baboon (256×256)** | G | 7.9974 | 7.9973 |
| | B | 7.9971 | 7.9972 |
| **House (256×256)** | R | 7.9997 | 7.9974 |
| | G | 7.9997 | 7.9977 |
| | B | 7.9996 | 7.9977 |
| | R | 7.9997 | 7.9993 |
| **Airplane (512×512)** | G | 7.9997 | 7.9993 |
| | B | 7.9996 | 7.9994 |
| **Pepper (512×512)** | R | 7.9995 | 7.9994 |
| | G | 7.9994 | 7.9993 |
| | B | 7.9993 | 7.9994 |

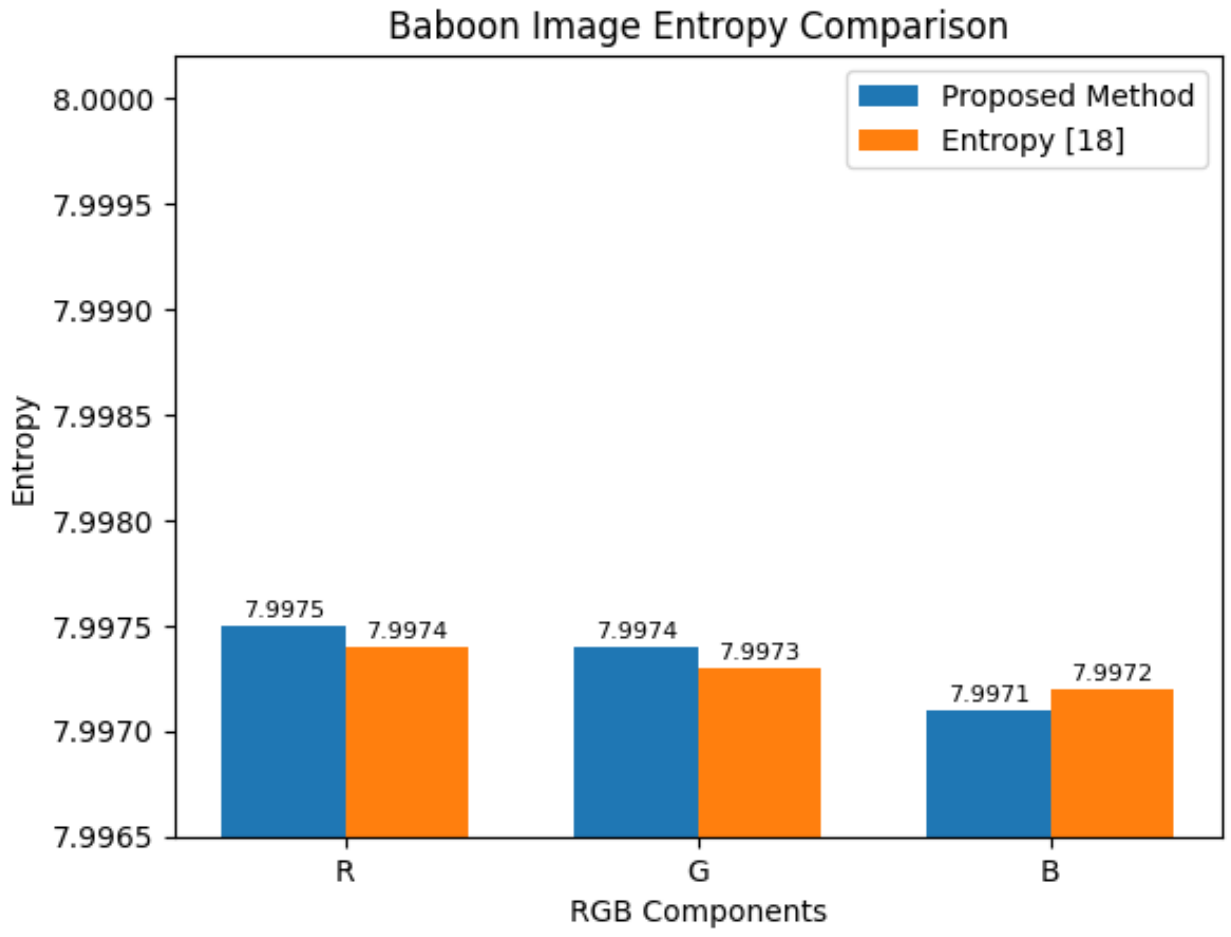

Figure 9. Baboon Image Entropy Comparison

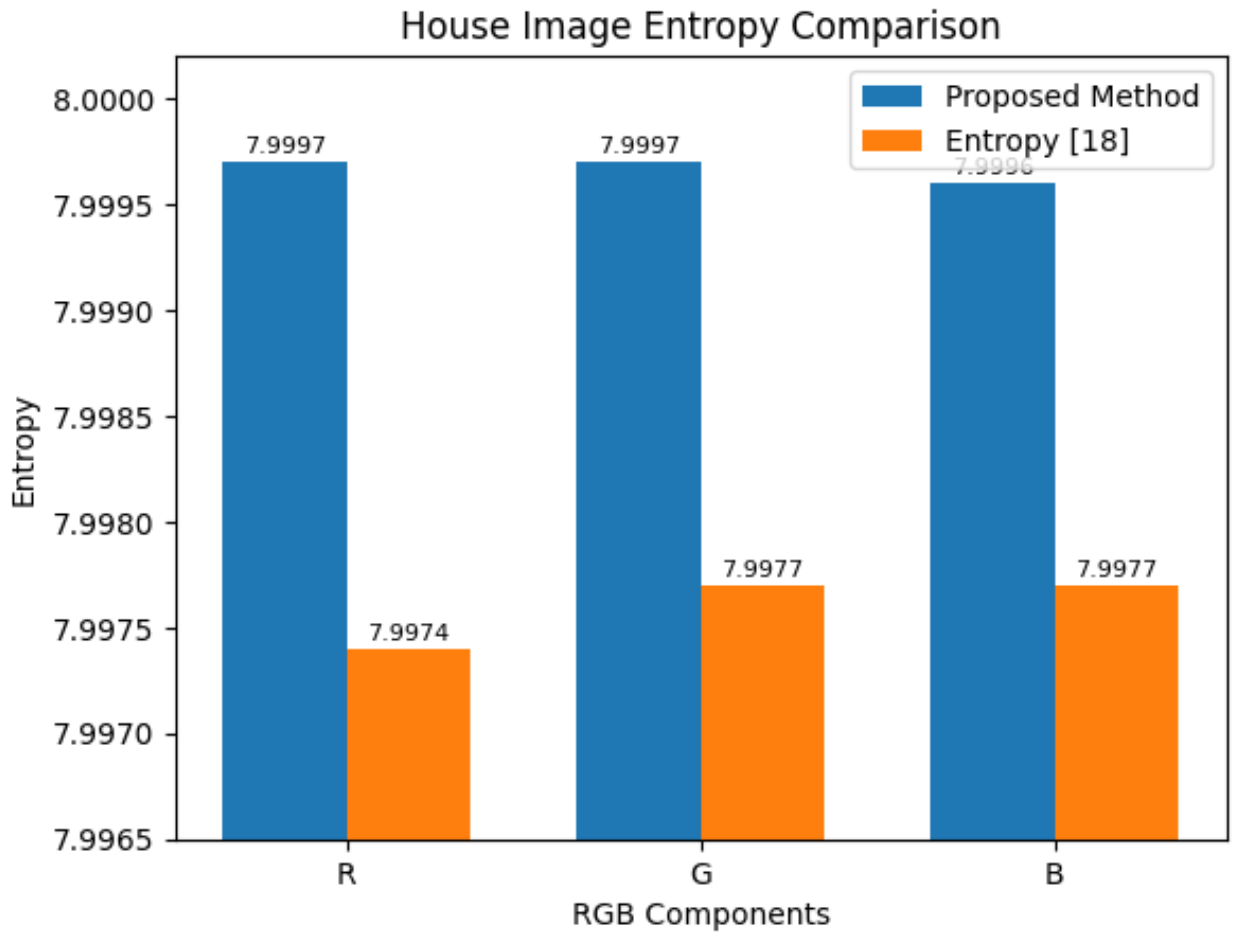

Figure 10. House Image Entropy Comparison

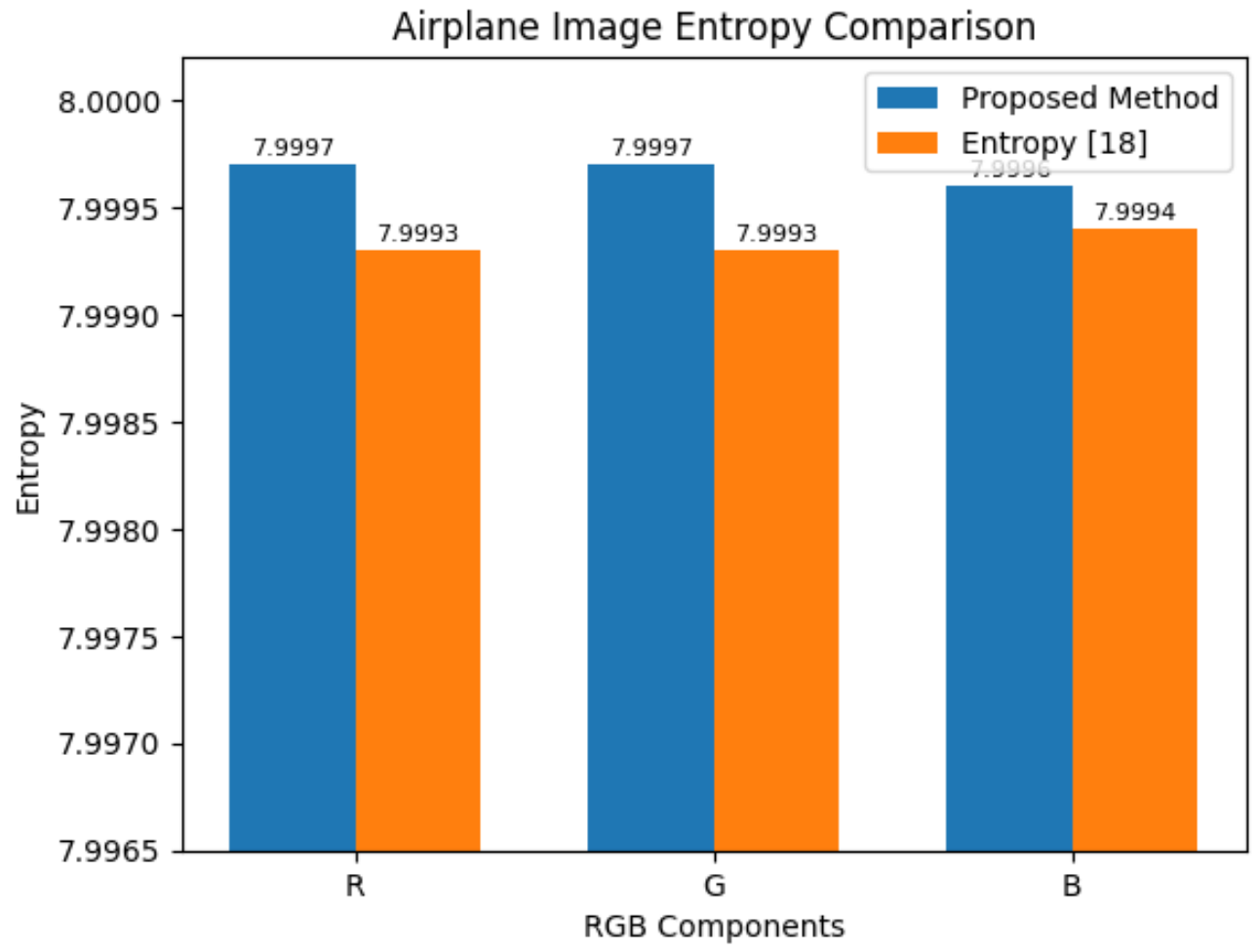

Figure 11. Airplane Image Entropy Comparison

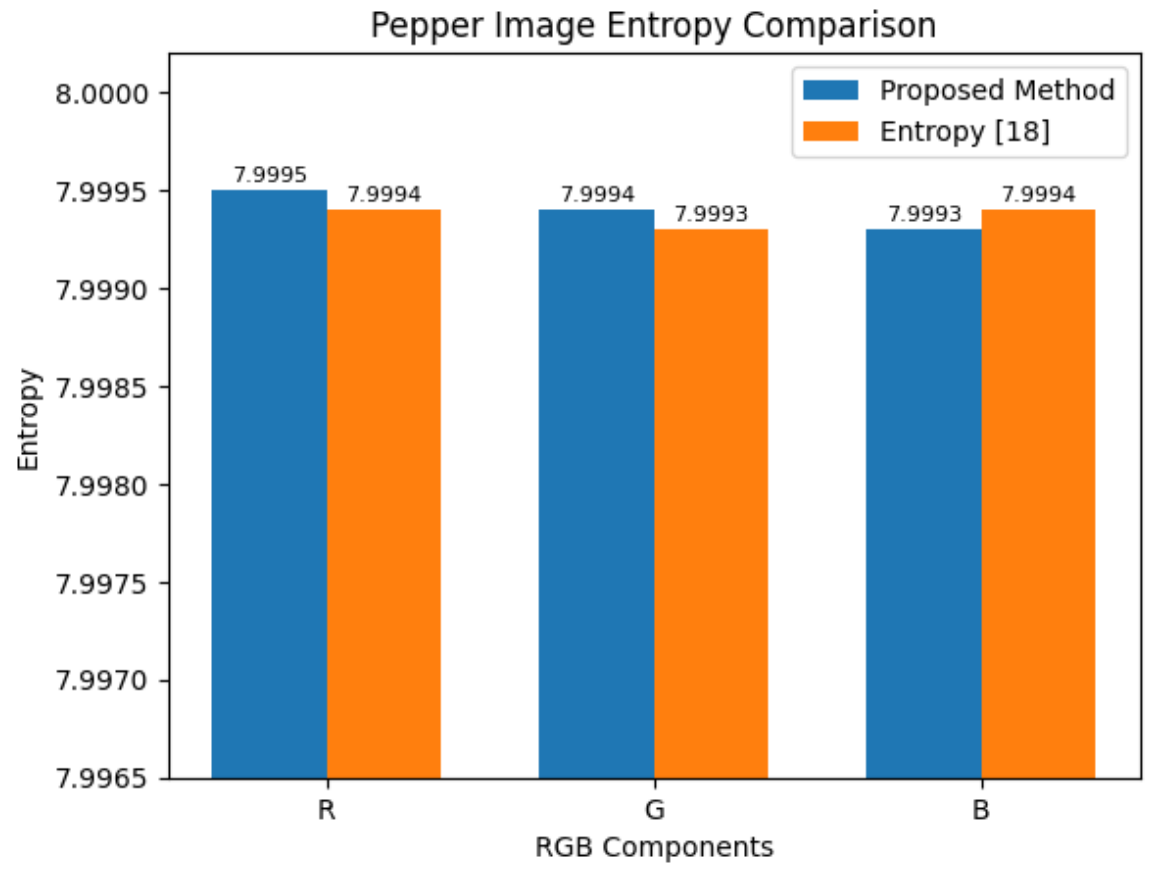

Figure 12. Pepper Image Entropy Comparison

The entropy values approach the ideal value of 8.0 for all image sizes, and the RGB components of the proposed CNN-based S-box encryption are very close. This is evidence of the efficiency of the algorithm in producing highly random-looking cipher images that are not susceptible to statistical analysis. The proposed method provides a slightly better entropy consistency compared to [18], particularly in larger images; thus, it is again suitable in applications that concern secure image encryption.

## 5 Conclusions And Future Work

This study suggested a new image encryption algorithm using Convolutional Neural Network (CNN) generated S-boxes and tested its performance. The technique utilizes the adaptive learning ability of CNNs to create very dynamic and complicated substitution boxes, which are the primary element of the encryption process. The extensive experimentation performed with benchmark images of different sizes sharpened the very high security and performance parameters of the proposed scheme. Entropy, NPCR, UACI, correlation coefficient, histogram uniformity, and computational time, which are key cryptographic measures, proved that the method is robust against statistical and differential attacks, and it encrypts and decrypts in real time. The CNN-based S-box method is more random and more resistant to cryptanalysis than the conventional encryption algorithms (such as XOR and AES). These encouraging findings make the suggested technique one of the possible and effective solutions to implement secure image transmission and storage in present multimedia and communication systems.

## Acknowledgement

The authors would like to show their heartfelt thanks to the University of Technology - Iraq, Middle Technical University (MTU) - Iraq, and the University of Tabriz - Iran in supporting the present research by way of granting the required academic resources and facilities.

## Data Availability

The dataset that was used in this study is the publicly available USC-SIPI Image Database that can be accessed by following this URL: http://sipi.usc.edu/database/.

These standard test images (Lena, Peppers, Baboon, etc) were applied to test the performance of the proposed CNN-based image encryption technique.

## Authors' Contributions

Fadhil Abbas Fadhil has developed the system encryption model, implemented the system, and prepared the manuscript. Mohammad-Reza Feizi-Derakhshi directed the research procedure.

Ans Ibrahim Mahameed and Nikolai Safiullin assisted in the emphasis of the thought, and also thoroughly reviewed the paper.

The experimental evaluation, verification of the references, and participation in the data analysis and revision of the final manuscript were done by Correspondence Author Maryam Mahdi Alhusseini. The final version of the manuscript was read by all authors, who have approved it.

## Conflict of Interest

The authors indicate that they have no conflict of interest as far as the publication of the current paper is concerned.